\documentclass{ws-ijmpd}

\begin{document}

\title{Brane World Dynamics and Adiabatic Matter creation}

\author{P Gopakumar}

\address{School of Pure and Applied Physics, Mahatma Gandhi University \\
P.D.Hills, Kottayam-686560, India.\\
gopan@iiap.res.in}

\author{GV Vijayagovindan}

\address{School of Pure and Applied Physics, Mahatma Gandhi University \\
P.D.Hills, Kottayam-686560, India.\\
gvvgovindan@rediffmail.com}

\maketitle

\begin{history}
\received{Day Month Year}
\revised{Day Month Year}
\comby{Managing Editor}
\end{history}

\begin{abstract}
We have treated the adiabatic matter creation process in various three-brane models by 
applying thermodynamics of open systems. The matter creation rate is found to affect the 
evolution of scale factor and energy density of the universe. We find modification at 
early stages of cosmic dynamics. In GB and RS brane worlds, by chosing appropriate 
parameters we obtain standard scenario, while the warped DGP model has different 
solutions. During later stages, since the matter creation is negligible 
the evolution reduces to FRW expansion, in RS and GB models.
\end{abstract}


\section{Matter Creation Scenario}	

Einstein's field equations are adiabatic and reversible and hence cannot provide an 
explanation for the origin of cosmological entropy.\cite{1}  But matter constituents may be produced 
quantum mechanically in the framework of Einstein's equations.\cite{2,3} The quantum fluctuations 
of the gravitational field near singularity, as well as the dynamics of any quantum field living on 
such a non stationary background, lead to matter- creating phenomena, which can in turn modify 
the behavior of the early universe.\cite{4} A detailed microscopic description of matter creation 
phenomena should be provided by a fully self-consistent quantum field theory on a curved space 
time.\cite{5} 

Many attempts have been made to treat the matter creation process at a 
phenomenological macroscopic level.\cite{6}\cdash\cite{13}  The basic idea is that matter 
creation contributes at the level of the Einstein field equations  as a negative pressure term. 
Prigogine et al. incorporated its effect in the classical Einstein field equations.\cite{1} At the expense 
of the gravitational field, matter creation can occur only as an irreversible process. The additional 
pressure term due to the creation depends on the creation rate. The total entropy within a 
comoving volume is not a constant, due to the variation of number of particles.  This approach 
explained the average effect induced on the dynamics of an isotropic expanding universe. 
In this work we study the effect of adiabatic matter creation on brane world models. We 
haven't considered the details of  the source of  created matter. Recently Tetradis has shown that 
absorption of energy flux from the bulk by the  three brane leads to acceleration or deceleration 
of  the universe.\cite{14} We have also derived a phantom field equivalent formulation.  This paper 
is organised as follows. In section 2 we give a brief description of RSII brane world model. In 
the section 3, thermodynamics of matter creation is explained. In section 4  we describe the 
effect of matter creation on brane worlds in general, and on RSII in section 5.  An equivalent 
phantom field formulation is given in section 6. In sections 7 and  8 we give the adiabatic matter 
creation effects on DGP and GB brane models.

\section{RSII Brane Model}

The idea that our 4-dimensional universe might be a 3-brane embedded in a higher dimensional 
space-time has attracted much attention over the last several years. Randall and Sundrum (RS) 
models\cite{15,16} are the most important. According to the brane world scenario, the physical 
fields in our 4-dimensional space-time are confined to the three brane, while gravity can freely 
propagate in the bulk space-time. We assume a Friedmann metric for 4-dimensional spacetime 
embedded in the bulk 5- dimensional space with a cosmological constant $\Lambda _5 $.

The Friedmann equation on the RSII brane is,\cite{17,18}

\begin{equation}
H^2  = \,\frac{{\kappa _4^2 }}{3}\rho \;(\,1 + \frac{\rho }{{2\sigma }}) + 
\frac{{\Lambda _4 }}{3} + \frac{C}{{R^4 }} - \frac{k}{{R^2 }}
\label{eqn1}
\end{equation}
with 

\begin{equation}
 \kappa _4^2  = \frac{{8\pi G}}{3},~~ and ~~ 
\frac{{\Lambda _4 }}{3} = \frac{{\sigma ^2 }}{{36}} + \frac{{\Lambda _5 }}{6},
\label{eqn2}
\end{equation}
where $ \Lambda _4 $  is the cosmological constant on the brane,  $\sigma $ 
is the brane tension, $ \rho $ is energy density of ordinary 
cosmological matter on the brane, k is the curvature index, and the term with C  (a constant) 
is the dark radiation term.\cite{19}  Cosmology constrains the amount of dark radiation to be at most 
10 percent of the energy density in photons.\cite{20}  In the following we will assume AdS bulk 
spacetime, and $C = 0,~k = 0$,~ and $\Lambda _4  = 0$

In RSII model the important change in Friedmann's equation compared to the usual four 
dimensional form is the appearance of a term proportional to $\rho $.
It tells us that if the matter energy density is much larger than the brane tension, i.e. 
$\rho >> \sigma $, the expansion rate H is proportional $\rho $. Only in the limit where the
 brane tension is much larger than the matter energy density, 
the usual behavior $H \propto \sqrt \rho$ is recovered.

\section{Thermodynamics of Matter Creation in Cosmology}

The Einstein's field equations with Bianchi identities give, for homogeneous and 
isotropic universes in closed systems, 

\begin{equation}
d(\rho V) =  - \tilde p\;dV.
\label{eqn3}
\end{equation}
where $\tilde p$ is the true thermodynamical pressure. 

This equation is used to describe 
an adiabatic evolution of a closed system with a comoving volume $V$. 
But in the presence of matter creation, the analysis is in the context of open systems.\cite{1,21} In 
this case the number of particles $N$ in a given volume is not fixed. 
For adiabatic transformation, 

\begin{equation}
d(\rho V) + p\;dV - \frac{h}{n}\,d(nV) = 0.
\label{eqn4}
\end{equation}
where $ n =\frac{N}{V} $ and  $ h = \rho  + p $ is the enthalpy per unit volume. 

In such a transformation, the heat received by the system is due 
to the change of the number of particles. In our cosmological context, this change 
is due to the 
transfer of energy from gravitation to matter. Correspondingly the entropy change $dS$,  for 
adiabatic transformation in open systems is,

\begin{equation}
T\;dS = \frac{h}{n}\,d(nV) - \mu \;(\rho V) = T\frac{s}{n}\;d(nV)
\label{eqn5}
\end{equation}
where $ \mu  = h - T\,s $ is the chemical potential and $s = \frac{S}{V} $. 
Therefore the only particle number variations admitted are such that

\begin{equation}
dN = d(nV) \ge 0
\label{eqn6}
\end{equation}
The Eq.~(\ref{eqn5}) can be expressed as,
\begin{equation}
\dot \rho  = \frac{h}{n}\,\dot n
\label{eqn7}
\end{equation}

\begin{equation}
p = \frac{{n\,\dot \rho  - \rho \,\dot n}}{{\dot n}}
\label{eqn8}
\end{equation}
where the over dot represents time derivative. The creation of matter corresponds to a 
supplementary pressure $p_c $, which is negative or zero depending on the presence 
or absence of particle production. Thus Eq.~(\ref{eqn5}) can be written as, 

\begin{equation}
d(\rho V) =  - (p + p_c )\,dV =  - \tilde p\;dV
\label{eqn9}
\end{equation}
where $p$ is the true thermodynamical pressure and 

\begin{equation}
p_c  =  - \frac{h}{n}\frac{{d(nV)}}{{dV}} =  - \frac{{\rho  + 
p}}{n}\;\frac{{d(nV)}}{{dV}}
\label{eqn10}
\end{equation}

This scenario is applied to cosmology as follows. 
In the case of an isotropic and homogeneous universe, we choose, for $V$, the value $V = R^3 
\left( \tau  \right)$  where $R$ is the scale factor. The above expression becomes,

\begin{equation}
p_c  =  - \frac{{\rho  + p}}{{3nH}}\left( {\dot n + 3nH} \right)
\label{eqn11}
\end{equation}
The conservation  equation becomes,
 \begin{equation}
 \dot \rho  =  - 3H( {\rho  + p + p_c } )  =  - 3H(\rho  + p) + 
 3H\frac{{(\rho  + p)}}{{3nH}}\,(\dot n + 3nH)  
\label{eqn12}
 \end{equation}
                       
\begin{equation}
\dot \rho  = \frac{{\dot n}}{n}\,(\rho  + p)
\label{eqn13}
\end{equation}
with     

 \begin{equation}
\frac{{\dot S}}{S} = \frac{{\dot N}}{N} = \frac{{\dot n + 3nH}}{n}
\label{eqn14}
\end{equation}
The balance equation for particle number density is,\cite{1} 

\begin{equation}
N^i  = n\,u^i ,\quad N^a _{;a}  = \dot n + 3nH = n\Gamma 
\label{eqn15}
\end{equation}
with  

\begin{equation}
N = n\,a^3, ~~ and ~~ \Gamma  = \frac{{\dot N}}{N} = \frac{\psi }{n}
\label{eqn16}
\end{equation}
where $ p $ is thermostatic pressure, $ u^i $ is  fluid 4-velocity, $n$ is particle number density, 
and $ \Gamma $ is matter creation rate. For $ \Gamma  > 0 $, there is particle production,
 and $ \Gamma  < 0 $ corresponds to annihilation. 

The above-mentioned macroscopic phenomenological approach is not completely self 
consistent since it cannot determine the expression for the rate of particle creation, which is an 
open degree of the scheme.

\section{Effect of Mater creation on Brane Worlds}

The existence of some kind of negative-pressure dark energy is usually assumed to 
explain the cosmic acceleration. Vacuum energy or a cosmological constant is the simplest dark 
energy candidate. It has been known that matter creation may lead to negative pressures.\cite{22}  
The modified Friedmann equation on RSII brane world indicates that the universe evolves 
slowly in the very early times. The process of matter creation may add to the accelerated 
expansion and thus can compensate to the effect of brane world at early times. A scenario with 
particle production driving cosmic acceleration was proposed by Zimdahl et al.\cite{23}  Using a 
phenomenological macroscopic description of matter creation,\cite{1} we investigate this scenario in 
the brane world context. However, we do not consider that particle creation as the only source 
for cosmic acceleration. It has been shown based on the supernova data that matter creation alone 
does not drive cosmic acceleration.\cite{24}  Our motivation is the following: the modified 
Friedmann equation on RSII brane world shows that there is transition for the universe evolution 
between an early high energy regime characterized by the behavior $R \propto t^{{1 
\mathord{\left/ {\vphantom {1 {3(1 + \omega )}}} \right.
 \kern-\nulldelimiterspace} {3(1 + \omega )}}} $
 and a late low energy regime characterized by the standard evolution $R \propto t^{{2 
\mathord{\left/ {\vphantom {2 {3(1 + \omega )}}} \right.
 \kern-\nulldelimiterspace} {3(1 + \omega )}}} $ .\cite{25,18}  The matter creation process due to
 its negative pressure can be the cause of this 
transition. Our attempt is to consider the possible effects on dynamics of the early universe due 
the matter creation.

\section{Effect of Adiabatic Matter Creation on the RSII Brane Model}

The field equations on the RS II brane model are,\cite{26}

\begin{equation}
H^2  = \frac{{R^2 }}{{R^2 }} = \frac{{\kappa ^2 }}{3}\rho \;(\,1 + \frac{\rho 
}{{2\sigma }}) + \frac{\Lambda }{3} + \frac{C}{{R^4 }} - \frac{k}{{R^2 }.}
\label{eqn17}
\end{equation}

\begin{equation}
\dot H = \frac{{\ddot R}}{R} - \frac{{\dot R^2 }}{{R^2 }} =  - \frac{{\kappa ^2 
}}{2}\rho \;(\,1 + \omega )\,\left( {1 + \frac{\rho }{\sigma }} \right) - \frac{{2C}}{{R^4 }} + 
\frac{\Lambda }{3}.
\label{eqn18}
\end{equation}
where $ \kappa ^2  = 8\pi G $ , and $ \omega  = \gamma  - 1 = \frac{p}{\rho } $ for perfect fluid. 
With particle creation, creation pressure, Eq.~(\ref{eqn11}), is added to thermostatic pressure $p$. 
\begin{equation}
p_c  = -\frac{{ \left( {\rho  + p} \right)}}{{3nH}}\,\psi 
\label{eqn19}
\end{equation}
For simplicity we take $Lambda = 0$. Also, since the dark radiation term decays rapidly 
we can put $ C = 0 $,  Eq.~(\ref{eqn18}) becomes,
\begin{equation}
\dot H = \frac{{\ddot R}}{R} - \frac{{\dot R^2 }}{{R^2 }} =  - \frac{{\kappa ^2 
}}{2}\rho \;(\,1 + \frac{{p + p_c }}{\rho })\,\left( {1 + \frac{\rho }{\sigma }} \right)
\label{eqn20}
\end{equation}

The energy conservation equation is unchanged in 5-d setup, which including particle creation, 
becomes, 

\begin{equation}
\dot \rho  + 3H\left( {\rho  + p + p_c } \right) = 0
\label{eqn21}
\end{equation}
For perfect fluid, using Eq.~(\ref{eqn19}) we get,
\begin{equation}
\dot \rho  + 3\frac{{\dot R}}{R}\gamma \rho \left( {1 - \frac{\psi }{{3nH}}} \right) = 0
\label{eqn22}
\end{equation}

We take the linear dependence of matter creation rate on Hubble parameter $H$ as,\cite{27} 

\begin{equation}
\psi  = \beta  {\kern 1pt} 3{\kern 1pt} n{\kern 1pt} H
\label{eqn23}
\end{equation}
where the parameter $\beta $ is defined in the interval $[0, 1]$ and assumed to be constant. 
The observational value of $H_0 t_0$ ($0.86 \le H_0 t_0  \le 1.91$) can be translated to 
the value of $\beta$. The joint result of SNe Ia and GRB  indicate that $\beta$ is 
($0.502 \pm 0.038$). The GRB data shows $\beta$ to be greater than $1/3$ 
(ie.,$ 0.349 \pm 0.185$), and the SNIa data alone shows ($0.537 \pm 0.040$).\cite{28}

This linear relation implies $\psi  \propto \rho $, at high energy densities. 
(Recently Freaza et. al. \cite{24} have considered a general form for 
the matter creation in non-brane scenario as,  
$\Psi  = 3\,\beta \,n\,H_0 \left( {{H \mathord{\left/
 {\vphantom {H {H_0 }}} \right.
 \kern-\nulldelimiterspace} {H_0 }}} \right)^\alpha  $, 
 where $\alpha $ and $\beta $ are dimensionless constants).
Now the solution of Eq.~(\ref{eqn22}) is,  

\begin{equation}
\rho  = \rho _0 \,R^{ - q} ~~  where ~~ q = 3\gamma \;(1 - \beta )
\label{eqn24}
\end{equation}
Substituting for $\rho $, and for $ C = k = \Lambda  = 0 $, Eq.~(\ref{eqn17}) becomes,

\begin{equation}
\frac{{\dot R^2 }}{{R^2 }} - \frac{{\kappa ^2 }}{3}\rho _0 \;R^{ - q}  - \frac{{\kappa 
^2 }}{{6\sigma }}\rho _0 ^2 \,\;R^{ - 2q}  = 0
\label{eqn25}
\end{equation}
Defining $ R^q  = X $, this becomes,  

\begin{equation}
\dot X^2  = \frac{{\kappa ^2 \,\rho _0 \;q^2 }}{{3\,\,}}\;\;X + \frac{{\kappa ^2 \;\rho _0 
^2 \;q^2 }}{{6\,\,\sigma }} = A\;X + B
\label{eqn26}
\end{equation}
Solving we get

\begin{equation}
t - t_0  = \frac{2}{{\sqrt A }}\sqrt {X + \frac{B}{A}} 
\label{eqn27}
\end{equation}
Using the boundary condition, $X = R^q  = 0$, when $ t = 0$ , we find the solution for 
the scale factor as follows,

\begin{equation}
R = \left( {\frac{{\kappa ^2 \;\rho _0 \;q^2 }}{{12}}\,t^2  + \frac{{\kappa \;\rho _0 
\;q}}{{\sqrt {6\,\sigma } }}\,t} \right)^{\frac{1}{q}} 
\label{eqn28}
\end{equation}

The FRW evolution laws are regained for $\beta  = \frac{1}{2}$, and  $
t <  < \frac{{12}}{{\kappa q\sqrt {6\sigma } }} $, which corresponds to high 
energy densities. Thus we get, for radiation 
$R \propto t ^{{\raise0.5ex\hbox{$\scriptstyle 1$}
\kern-0.1em/\kern-0.15em
\lower0.25ex\hbox{$\scriptstyle 2$}}} $, and  
for dust  $R \propto t ^{{\raise0.5ex\hbox{$\scriptstyle 2$}
\kern-0.1em/\kern-0.15em \lower0.25ex\hbox{$\scriptstyle 3$}}} $. 

Using the phenomenological limit, $\sigma  \ge 1\left( {MeV} \right)^4 $
,\cite{29} the inequality, $
t <  < \frac{{12}}{{\kappa q\sqrt {6\sigma } }} $ gives $t <  < 10^{30} s$. 
This means that our model is valid for any time in the evolution of the universe. The 
value of $q$ depends on $\beta$. Assuming that the rate of matter creation is insignificant 
during later stages of the evolution of the universe, the evolution of scale factor
 reduces to the FRW relation.

\section{Scalar Field Dynamics}

The effect of matter creation may be translated into the dynamics of a minimally coupled 
scalar field. The effect of negative pressure due to the matter creation is equivalent
 to a suitable scalar field potential. Simple models of a super accelerated universe 
include a phantom field, i.e., a minimally coupled scalar field with negative kinetic 
energy.\cite{30}\cdash\cite{32}  It has also been proposed that early inflation and 
late time acceleration of the universe can be unified in a single theory based on a
 phantom field.\cite{33} Phantom type of matter may also arise from a bulk viscous stress 
due to particle production.\cite{34}
 
We consider particle production from a phantom field, and derive the corresponding 
potential. The field equation including the phantom field, $C$ is,

\begin{equation}
R_{\mu \nu }  - \frac{1}{2}R{\kern 1pt} g_{\mu \nu }  = 8\pi G\left( {\left( {\rho _m  + 
P_m } \right)U_\mu  U_\nu   + Pg_{\mu \nu }  - \partial _\mu  C\,\partial _\nu  C + 
\frac{1}{2}{\kern 1pt} g_{\mu \nu } g^{\alpha \beta } \partial _\alpha  C\,\partial _\beta  C} 
\right)
\label{eqn29}
\end{equation}

The RHS can be written as, $8\pi G\left( {T_{_{\mu \nu } }^{fluid}  + T_{_{\mu \nu } 
}^{phantom} } \right)$, then the sum 
$T_{\mu \nu }  = T_{_{\mu \nu } }^{fluid}  + T_{_{\mu \nu } }^{phantom} $
 is conserved by the Bianchi identity but the individual terms are not. This was 
interpreted in the Steady State theory\cite{35} as creation of matter by the 
creation field in order to maintain a constant matter density.  

The conservation equation, assuming Friedmann metric, is,
\begin{equation}
\dot \rho  + 3H(\rho  + p) = 0
\label{eqn30}
\end{equation}

We assume the total energy and pressure consist of matter and phantom field components as,
\begin{equation}
\rho  = \rho _m  + \rho _\phi  
\label{eqn31}
\end{equation}

\begin{equation}
\dot \rho _m  + \dot \rho _\phi   + 3H(\rho _m  + \rho _\phi   + p_m  + p_\phi  ) = 0
\label{eqn32}
\end{equation}

\begin{equation}
\dot \rho _m  + 3H\left( {\rho _m  + p_m  + \frac{1}{{3H}}\dot \rho _\phi   + \rho _\phi   
+ p_\phi  } \right) = 0
\label{eqn33}
\end{equation}

To obtain a phenomenological classical description of the essentially quantum creation process, 
we assume that the creation pressure is proportional to the energy density and pressure of the 
scalar field as,

\begin{equation}
p_c  = \frac{1}{{3H}}\dot \rho _\phi   + \rho _\phi   + p_\phi  
\label{eqn34}
\end{equation}
Now the Eq.~(\ref{eqn33}) becomes, 

\begin{equation}
\dot \rho _m  + 3H\left( {\rho _m  + p_m  + p_c } \right) = 0
\label{eqn35}
\end{equation}
The expression for creation pressure, from Eq.~(\ref{eqn11}) is, 

\begin{equation}
p_c  =  - (\rho _m  + p_m )\frac{\psi }{{3nH}} =  - (\rho _m  + p_m )\beta 
\label{eqn36}
\end{equation}
For perfect fluid matter, $p_m  = (\gamma  - 1)\rho _m $

\begin{equation}
\dot \rho _m  + 3\,\frac{{\dot R}}{R}\,\gamma \, \rho _m \left( {1 - \beta } \right) = 0
\label{eqn37}
\end{equation}
The solution is, 
\begin{equation}
\rho _m  = \frac{{\rho _{m0} }}{{R^q }}
\label{eqn38}
\end{equation}
where  $q = 3\gamma \left( {1 - \beta } \right)$
For phantom fields,\cite{36,30} the energy density, 
\begin{equation}
\rho _\phi   =  - \frac{1}{2}\dot \phi ^2  + V(\phi )
\label{eqn39}
\end{equation}
and pressure, 
\begin{equation}
p_\phi   =  - \frac{1}{2}\dot \phi ^2  - V(\phi ),
\label{eqn40}
\end{equation}
where $V$ is the potential of phantom field.
The scalar field equations are,
\begin{equation}
\frac{\partial }{{\partial x^i }}(\sqrt { - g} \,g^{ij} \frac{{\partial \phi }}{{\partial x^i 
}}) = 0
\label{eqn41}
\end{equation}

\begin{equation}
\frac{\partial }{{\partial t}}\,(R^3 \dot \phi ) = 0
\label{eqn42}
\end{equation}
The solution is, 
\begin{equation}
\dot \phi  = \frac{{\phi _0 }}{{R^3 }}
\label{eqn43}
\end{equation}
Using this solution and Eqs.~(\ref{eqn24}), (\ref{eqn25}) and (\ref{eqn26}) in Eq.~(\ref{eqn23}) we get the potential as,
\begin{equation}
V = \frac{{3\gamma \rho _{m0} \beta }}{q}\frac{1}{{R^q }} = \frac{{3\gamma \rho 
_{m0} \beta }}{q}\left( {\frac{{\dot \phi }}{{\phi _0 }}} \right)^{q/3} 
\label{eqn44}
\end{equation}
The modified Friedmann equation, Eq.~(\ref{eqn1}), for very high energy densities becomes,

\begin{equation}
H^2  = \frac{{\dot R^2 }}{{R^2 }} = \frac{{\kappa ^2 }}{{6\sigma }}\rho ^2 
\label{eqn45}
\end{equation}
Using Eq.~(\ref{eqn31}) for $\rho$,

\begin{equation}
\frac{{\dot R}}{R} = \frac{{ \pm \kappa }}{{\sqrt {6\sigma } }}\left( {\rho _m  + \rho 
_\phi  } \right)
\label{eqn46}
\end{equation}
We can write this equation as, 
\begin{equation}
\frac{{\dot R}}{R} - \frac{A}{{R^q }} + \frac{B}{{R^6 }} = 0
\label{eqn47}
\end{equation}
where  
\begin{equation}
A = \frac{{\kappa \rho _{m0} }}{{\sqrt {6\sigma } }}\left( {1 + \frac{{3\gamma \beta }}{q}} 
\right) = \frac{{3\kappa \rho _{m0} \gamma }}{{q\sqrt {6\sigma } }}
~, ~~  
B = \frac{{\kappa \phi _0^2 }}{{2\sqrt {6\sigma } }}
\label{eqn48}
\end{equation}
Defining, $ R^q  = X $ where $q = 3\gamma (1 - \beta )$, the above equation can be 
integrated as,

\begin{equation}
\int {\frac{{dX}}{{Aq - BqX^{1 - 6/q} }}}  = \int {dt} 
\label{eqn49}
\end{equation}
						
This integral  can be solved for a given value of q.
With $\beta  = {1 \mathord{\left/ {\vphantom {1 2}} \right.
 \kern-\nulldelimiterspace} 2}$, (the value which gives the FRW law for
 the RSII case), and $\gamma  = {4 \mathord{\left/
 {\vphantom {4 3}} \right.
 \kern-\nulldelimiterspace} 3}$
 (radiation), $q$ becomes $2$, thus the solution of Eq.~(\ref{eqn49}) is,

\begin{equation}
t - t_0  = \frac{X}{{A\,q}} - \frac{{Bq\,\arctan h\left( {\sqrt {\frac{A}{B}} X} 
\right)}}{{\left( {A\,q} \right)^{{3 \mathord{\left/
 {\vphantom {3 2}} \right.
 \kern-\nulldelimiterspace} 2}} \sqrt B }}
\label{eqn50}
\end{equation}
Neglecting the second term in comparison with the first, and  $t_0 = 0$ we get,

\begin{equation}
R = 2A\,\,t^{{1 \mathord{\left/
 {\vphantom {1 2}} \right.
 \kern-\nulldelimiterspace} 2}}  = \frac{{3\kappa \rho _{m0} \gamma }}{{\sqrt {6\sigma 
} }}\,\,t^{{1 \mathord{\left/ {\vphantom {1 2}} \right.
 \kern-\nulldelimiterspace} 2}} 
\label{eqn51}
\end{equation}
Substituting in Eq.~(\ref{eqn43}) and solving for $\phi$ we get,

\begin{equation}
t = \frac{{4\phi _0^2 \left( {6 \,\sigma } \right)^3 }}{{\left( {3 \,\kappa \, \rho _{m0} \gamma 
} \right)^6 }}\frac{1}{{\phi ^2 }}
\label{eqn52}
\end{equation}
Using Eqs.~(\ref{eqn44}), (\ref{eqn51}) and (\ref{eqn52}), we get the expression for the potential in terms of 
phantom field as,

\begin{equation}
V = \frac{{9\beta \kappa ^4 (\rho _{m0} \gamma )^5 }}{{32\sigma ^2 \phi _0^2 }}\phi ^2 
\label{eqn53}
\end{equation}
							
This potential of the phantom field gives the same evolution for the sale factor as 
produced by the adiabatic matter creation, in the high density regime. In this sense the 
phantom field model is equivalent to the adiabatic matter creation model discussed 
previously in section 4.

\section{Gauss-Bonnet brane model}

Gauss-Bonnet (GB) combination arises as the leading order for quantum corrections in 
the heterotic string effective action. In five dimensions it represents the unique 
combination of 
curvature invariants that leads to second-order field equations, linear in the second 
derivatives in 
the metric tensor, and it is ghost-free.\cite{37}\cdash\cite{44}  The GB correction to the 
Einstein-Hilbert 
action will disrupt mildly the degeneracy between tensor and scalar perturbations.\cite{45} 
Recently there is a renewed interest upon this scenario. 
Here we consider the effect of adiabatic matter creation in this scenario.

In the RSII scenario with a GB term, the effective Friedmann equation describing the 
motion of brane containing general perfect fluid may be derived from a generalization of the 
Birkhoff's theorem.\cite{46} The 5-dimensional bulk action for the GB braneworld scenario is given 
by,\cite{47}

\begin{equation}
S = \frac{1}{{2\kappa _5^2 }}\int_M {d^5 x\sqrt { - g} \left( {R - 2\Lambda ^{(5)}  +
 \alpha \left[ {GB} \right]} \right)}  + S_{\partial M}  + S_m 
\label{eqn54}
\end{equation}
where $\left[ {GB} \right] = \left( {R^2  - 4\,R_{AB} \,R^{AB}  + R_{ABCD} R^{ABCD} } \right)
$ is the Gauss-Bonnet term, $\alpha  \ge 0 $ with dimension $(mass)^2 $ is the Gauss-Bonnet coupling, $
\Lambda ^{(5)}  < 0 $  is the bulk cosmological constant and   
$\kappa _5^2  = 8\pi M_5^{ - 3} $
determines the 5d Planck scale. The $S_{\partial M} $
is the boundary term that is required to cancel normal derivatives of the metric tensor which 
arises when varying the action with respect to the metric, \cite{48} and $S_m $
 is the action for matter confined on the brane. We assume a  Z2 symmetry across the brane. We 
will consider the case that a perfect fluid matter source with density $\rho$ is confined 
to a brane with tension $\sigma $.

A new constant is defined, 
$b = \left( {1 + \frac{4}{3}\alpha \Lambda ^{(5)} } \right)^{{3 \mathord{\left/
 {\vphantom {3 2}} \right.
 \kern-\nulldelimiterspace} 2}} $. Since $\Lambda ^{(5)}  < 0 $, there is an upper 
bound on $\alpha$, $
\alpha  \le \frac{3}{{4\,\left| {\Lambda ^{(5)} } \right|}} \equiv \alpha _u $.
In order that the standard Friedmann equation be recovered at sufficiently low energy 
scales ($ \rho < < \sigma$), $
\kappa _4^2  \equiv \frac{{8\pi }}{{M_4^2 }} = \frac{{\kappa _5^2 \,\sigma 
}}{{6\,b^{{2 \mathord{\left/
 {\vphantom {2 3}} \right.
 \kern-\nulldelimiterspace} 3}} }} $, where $M_4$ is the 4d Planck scale.
Introducing a dimensionless variable $\chi$, the modified Friedmann equations are

\begin{equation}
\rho  + \lambda  = \left( {\frac{{\sigma \, b^{{1 \mathord{\left/
 {\vphantom {1 3}} \right.
 \kern-\nulldelimiterspace} 3}} }}{{3 \,\alpha \, \kappa _4^2 }}} \right)^{{1 \mathord{\left/
 {\vphantom {1 2}} \right.
 \kern-\nulldelimiterspace} 2}} \sinh {\chi}
\label{eqn55}
\end{equation}
and,     

\begin{equation}
H^2  = \frac{1}{{4\alpha }}\left[ {b^{{1 \mathord{\left/
 {\vphantom {1 3}} \right.
 \kern-\nulldelimiterspace} 3}} \,\cosh \left( {\frac{{2 \chi}}{3}} \right) - 1} \right]
\label{eqn56}
\end{equation}
where $\sigma$ is the brane tension.
In the high energy approximation, $\rho  >  > \sigma $, we can write it as, 

\begin{equation}
H^2  = A\rho ^{2/3} 
\label{eqn57}
\end{equation}
In order for that the 4d effective cosmological constant vanishes, the brane tension 
should satisfy
$ \sigma  = \frac{3}{2}\frac{{1 - b^{{1 \mathord{\left/
 {\vphantom {1 3}} \right.
 \kern-\nulldelimiterspace} 3}} }}{{\alpha \kappa _4^2 }} $
. We can see that  $
\sigma  \approx {{\left| {\Lambda ^{(5)} } \right|} \mathord{\left/
 {\vphantom {{\left| {\Lambda ^{(5)} } \right|} {\kappa _4^2 }}} \right.
 \kern-\nulldelimiterspace} {\kappa _4^2 }} $
for any  $ 0 \le \alpha  \le \alpha _u $.

\subsection{Mater Creation on GB Model}

Following a similar calculation as in RS II model in section 5, with a substitution $X = 
R^{\frac{2}{3}q} $ in Eq.~(\ref{eqn57}), we get
\begin{equation}
\dot X^2  = \frac{{4q^2 \rho _0 ^{2/3} A}}{9}X
\label{eqn58}
\end{equation}
The solution with the boundary condition, $X = R^q  = 0 $ when $ t = 0 $, gives

\begin{equation}
t = 2\sqrt {\frac{X}{A}} 
\label{eqn59}
\end{equation}
                     
\begin{equation}
R = \left( {\frac{{q^2 \rho _0 ^{2/3} A}}{9}} \right)^{\frac{3}{2}\,\,q} 
\;t^{\frac{3}{q}} 
\label{eqn60}
\end{equation}
where $ q = 3\gamma \,\left( {1 - \frac{\psi }{{3nH}}} \right) $
 and  $ \beta  = \frac{\psi }{{3nH}}$.

This evolution is for high energy densities $ t <  < 1 $ , since we considered the
 modified Friedmann equation for that regime.  The case of $ \beta  = {1 \mathord{\left/
 {\vphantom {1 2}} \right. \kern-\nulldelimiterspace} 2}$ , which was considered 
in RSII model, gives for radiation and dust as $ R \propto t^{{3 \mathord{\left/
 {\vphantom {3 2}} \right. \kern-\nulldelimiterspace} 2}} $ and $ R \propto t^2 $
respectively. These are modified evolution equations compared to standard cases without matter 
creation.

\section{DGP Brane Model}

An alternative scenario to the RS II brane model was proposed by Dvali, Gabadadze and 
Porrati (DGP) .\cite{49}\cdash\cite{51}  In this model the extra dimension is infinitely large. The key feature
 is the 
presence of a four-dimensional curvature scalar in the action. The DGP model predicts that 4D 
Newtonian gravity on a brane world is regained at distances shorter than a given crossover
 scale 
$r_c$ (high energy limit), whereas 5D effects become manifest above that scale 
(low energy limit). 
The interesting feature is that the effective cosmological constant on the brane
 can be extremely 
reduced in contrast to the case of the Randall-Sundrum model even if a bulk cosmological 
constant and a brane tension are not fine-tuned. Also, the model can explain late-time 
acceleration without having to invoke a cosmological constant or quintessential matter.\cite{52} 
Here we consider a generalized DGP brane model of Sahni and Shtanov.\cite{53}  

The action of the generalized DGP model is,\cite{53}

\begin{equation}
S = S_{bulk}  + S_{brane} 
\label{eqn61}
\end{equation}
where, 

\begin{equation}
S_{bulk}  = \int\limits_ {d^5 } X\,\sqrt { - g^{(5)} } \left[ {\frac{1}{{2\kappa _5^2 
}}R{}^{(5)} + L_m^{(5)} } \right],~~and,~~
\label{eqn62}
\end{equation}

\begin{equation}
S_{brane}  = \int\limits_M {d^4 } x\,\sqrt { - g} \left[ {\frac{1}{{\kappa _5^2 }}K^ 
\pm   + L_{brane} \left( {g_{\alpha \beta } ,\chi } \right)} \right]
\label{eqn63}
\end{equation}

Here $\kappa _5^2 $ is the 5-dimensional gravitational constant, $R{}^{(5)}$
 and $L_m^{(5)} $ are the 5-dimensional curvature scalar and the matter Lagrangian 
in the bulk, respectively. $x^\mu  \left( {\mu  = 0,1,2,3} \right)$
 are the induced 4-dimensional coordinates on the brane, $K^ \pm  $
 is the trace of extrinsic curvature on either side of the brane and 
$L_{brane} \left( {g_{\alpha \beta } ,\chi } \right)$
 is the effective 4-dimensional Lagrangian, which is given by a generic functional 
of the brane metric $g_{\alpha \beta } $ and matter fields $\chi $ on the brane.

The brane Lagrangian consists of the following terms,

\begin{equation}
L_{brane}  = \frac{1}{{2\kappa _4^2 }}R - \sigma  + L_m 
\label{eqn64}
\end{equation}
where $ \sigma $ is the brane tension.

The 5-D bulk space includes only a cosmological constant $\Lambda {}^{\left( 5 \right)}$
. It is a generalized version of Dvali-Gabadadze-Porrati model, which is obtained by setting 
$ \sigma  = 0 $ as well as $ \Lambda {}^{\left( 5 \right)} = 0 $. From the field equation
 induced on the brane, the effective 4-dimensional cosmological constant 
on the brane is, $ \Lambda  = \frac{1}{2}\left( {\Lambda {}^{\left( 5 \right)} 
+ \frac{1}{6}\kappa _5^4 } \right) $.

Considering a Friedmann-Robertson-Walker metric on the brane, the Friedmann equation is 
found to be,\cite{52}

\begin{equation}
H^2  + \frac{k}{{R^2 }} = \frac{{\kappa _4^2 }}{3}\left[ {\rho  + \rho _0 \left( {1 + 
\delta {\cal A}\left( {\rho ,R} \right)} \right)} \right]
\label{eqn65}
\end{equation}
where $ \delta $  is either +1 or -1, and $ {\cal A}$ is given by,

\begin{equation}
{\cal A} = \left[ {{\cal A}_0^2  + \frac{{2\eta }}{{\rho _0 }}\left( {\rho  - 
\frac{{\varepsilon _0 }}{{\kappa _4^2 R^4 }}} \right)} \right]^{\frac{1}{2}} 
\label{eqn66}
\end{equation}
with   ${\cal A}_0  = \sqrt {1 - 2\eta \frac{\Lambda }{{\kappa _4^2 \rho _0 }}} $,
$ \eta  = \frac{{6m_5^6 \kappa _4^2 }}{{\rho _0 \,}}\quad \left( {0 < \eta  \le 1} \right)$
, and $\rho _0  = m_\sigma^4  + 6m_5^6 \kappa _4^2 $. There are two mass scales, $
m_\sigma   = \sigma ^{{1 \mathord{\left/
 {\vphantom {1 4}} \right.
 \kern-\nulldelimiterspace} 4}} $ and $m_5  = \kappa _5^{ - {2 \mathord{\left/
 {\vphantom {2 3}} \right. \kern-\nulldelimiterspace} 3}} $, in this model. 

At high energy limit, $\rho  >  > \rho _0 $, the modified Friedmann 
equation Eq.~(\ref{eqn65}) reduces to 

\begin{equation}
H^2  + \frac{k}{{R^2 }} \approx \frac{{\kappa _4^2 }}{3}\left( {\rho  + \delta \,\rho _0 
\,\sqrt {\,\,\frac{{2\eta }}{{\rho _0 }}\,\left( {\rho  - \frac{{\varepsilon _0 }}{{\kappa _4^2 R^4 
}}} \right)} } \right)
\label{eqn67}
\end{equation}
For simplicity, the curvature term and dark radiation term can be omitted. 
Then the Eq.~(\ref{eqn67}) can be further simplified to

\begin{equation}
H^2  = \frac{{\kappa _4^2 }}{3}\left( {\rho  + \delta \sqrt {2\eta \,\,\rho _0 \,\rho } } 
\right).
\label{eqn68}
\end{equation}
We will use this evolution law to investigate the effects of adiabatic matter creation.

\subsection{Matter Creation on DGP Model}

We assume the expression , $\psi  = \beta \,3n\,H $, for matter creation.
Defining, $ q = 3\gamma \;(1 - \beta )$, and inserting the solution, 
$ \rho  = \rho _0 \,R^{ - q} $, of the conservation equation 
(with modified creation pressure term) in Eq.~(\ref{eqn68}), we get

\begin{equation}
\frac{{\dot R^2 }}{{R^2 }} - \frac{{\rho _0 \kappa _4^2 }}{3}\;R^{ - q}  - \delta 
\frac{{\kappa _4^2 \sqrt {2\eta \rho _0 } }}{3}\,\;R^{ - q/2}  = 0
\label{eqn69}
\end{equation}
Defining $ R^{q/2}  = X $, we get 

\begin{equation}
\frac{{\dot R}}{R} = \frac{{\dot X}}{{X\;q/2}}
\label{eqn70}
\end{equation}
Substituting this in the above equation,

\begin{equation}
\frac{{\dot X^2 }}{{X^2 \,{{q^2 } \mathord{\left/
 {\vphantom {{q^2 } 4}} \right.
 \kern-\nulldelimiterspace} 4}}} - \frac{{\kappa _4^2 \rho _0 }}{3}\;\frac{1}{{X^2 }} - 
\delta \frac{{\kappa _4^2 \sqrt {2\eta \rho _0 } }}{3}\,\;\frac{1}{X} = 0
\label{eqn71}
\end{equation}
This equation can be written as, 

\begin{equation}
\dot X^2  = AX + B
\label{eqn72}
\end{equation}
Where $A = \frac{{\delta \,\kappa _4^2 \sqrt {2\eta \rho _0 } }}{3}\frac{{q^2 }}{4}$
 and $ B = \frac{{\kappa _4^2 \rho _0 }}{3}\frac{{q^2 }}{4}$. The solution of Eq.~(\ref{eqn72}) is

\begin{equation}
t - t_0  = \frac{2}{{\sqrt A }}\sqrt {X + \frac{B}{A}} 
\label{eqn73}
\end{equation}
With the boundary condition, $X = R^q = 0$, when $ t = 0$, the constant $t_0$ becomes
$ - \frac{{2\;\sqrt B }}{A}$. 
Therefore we get,

\begin{equation}
R = \left( {\delta \frac{{q^2 \sqrt {2\eta \rho _0 } }}{{12\mu ^2 }}t^2  + 
\frac{{\;q}}{2}\sqrt {\frac{{\rho _0 }}{{3\mu ^2 }}} \,\,\,t} \right)^{\frac{2}{q}} 
\label{eqn74}
\end{equation}
where   $q = 3\gamma \,\left( {1 - \frac{\psi }{{3nH}}} \right)$,  $\delta  =  \pm 1$
 and  $ \beta  = \frac{\psi }{{3nH}}$.

For $\beta  = \frac{1}{2}$ (special solution in the case of RSII), and for $
t <  < \frac{{\sqrt 6 \,\mu }}{{\,q\,\eta }} $
, which corresponds to high energy densities, we get, 

\begin{equation}
for~ radiation,~ R \propto t 
\end{equation}

\begin{equation}
for~~ dust, ~~~R \propto t^{{4 \mathord{\left/ {\vphantom {4 3}} \right.
 \kern-\nulldelimiterspace} 3}}. 
\label{eqn75}
\end{equation}
These are new evolution equations different from FRW laws. 
For $t >  > \frac{{\sqrt 6 \,\mu }}{{\,q\,\eta }}$ (later stage) the rate of 
matter creation is assumed to be insignificant, thus the evolution of scale 
factor is unaffected.

\section{Discussion}
 
The adiabatic matter creation rate is found to affect the evolution of scale factor and 
energy density of the universe, in all the cases we have considered. The modified Friedmann 
equation on brane world shows that the universe evolves slowly in the very early times. Thus 
there is a transition of the universe evolution between an early high-energy regime and a 
late low 
energy regime. The negative pressure due to the matter creation process can explain this 
transition. For the RSII model the standard FRW evolution law can be obtained for a matter 
creation rate $\beta  = {1 \mathord{\left/ {\vphantom {1 2}} \right.
 \kern-\nulldelimiterspace} 2}$. This value is consistent with the observed data. 
We have derived a scalar field 
equivalent formulation using a phantom field. A potential is derived which would give the 
dynamics same as that produced by adiabatic matter creation. In the case of Gauss-Bonnet and warped 
DGP models there are modifications in evolution of the universe with matter creation. For later 
stages of the evolution of the universe, the rate of matter creation is insignificant, thus the 
evolution of scale factor is unchanged.

\section*{Acknowledgement}

PG thanks CSIR, India for a research fellowship, and Dr.K.Indulekha, SPAP, for suggestions. 
We thank Prof. Naresh Dadhich, IUCAA for suggestions, and IUCAA, Pune for the hospitality 
during this work.

\end{document}